\documentclass[useAMS,twocolumn,usenatbib]{mn2e}

\def\OutputDriver{dvips}

\usepackage[\OutputDriver]{graphicx}
\usepackage{amssymb}
\usepackage{amsmath}
\usepackage{bm}

\newcommand{\be}{\begin{eqnarray}}
\newcommand{\ee}{\end{eqnarray}}
\newcommand{\bse}{\begin{subequations}}
\newcommand{\ese}{\end{subequations}}

\newcommand{\mrm}[1]{\mathrm{#1}}

\newcommand{\diff}[2][]{\mrm{d}^{#1}{#2}\,}

\newcommand{\km}{\ensuremath{\mathrm{km}}}

\newcommand{\kpc}{\ensuremath{\mathrm{kpc}}}
\newcommand{\Mpc}{\ensuremath{\mathrm{Mpc}}}

\newcommand{\Msolar}{\ensuremath{\mathrm{M}_\odot}}

\newcommand{\arcsect}{\ensuremath{\mathrm{arcsec}}}

\newcommand{\zS}{z_\mrm{S}}
\newcommand{\zL}{z_\mrm{L}}

\newcommand{\chiL}{\chi_\mrm{L}}

\def\21cm{\mbox{21-cm}}

\newcommand{\Mcluster}{M^\text{crit}_{200}}
\newcommand{\pixel}{\mathrm{p}}
\newcommand{\thetaE}{\theta_\mathrm{E}}

\begin{document}

\title[Cluster strong lensing in the Millennium Simulation]{
Cluster strong lensing in the Millennium Simulation: The effect of galaxies and structures along the line-of-sight
}
\author[Ewald Puchwein \& Stefan Hilbert]{
Ewald Puchwein$^1$
and
Stefan Hilbert$^{2,1}$
\\$^1$Max-Planck-Institut
f{\"u}r Astrophysik, Karl-Schwarzschild-Stra{\ss}e 1, 85741 Garching,
Germany
\\$^2$Argelander-Institut f{\"u}r Astronomie, Auf dem H{\"u}gel 71,
53121 Bonn, Germany
}
\date{\today}
\maketitle

\begin{abstract}
We use ray-tracing through the Millennium simulation to study how secondary matter
structures along the line-of-sight and the stellar mass in galaxies affect strong cluster lensing, in particular the cross-section for giant arcs.
Furthermore, we investigate the distribution of the cluster Einstein
radii and the radial distribution of giant arcs. 
We find that additional structures along the line-of-sight increase
the strong-lensing optical depth by $\sim10-25\%$, while strong-lensing
cross-sections of individual clusters are frequently boosted by as much
as $\sim50\%$.
The enhancement is mainly due to structures that are not correlated with
the lens.
Cluster galaxies increase the strong-lensing optical depth by
up to a factor of 2, while interloping galaxies are not significant.
We conclude that these effects need to be taken into account for
predictions of the giant arc abundance, but they are not large enough to
fully account for the reported discrepancy between predicted and
observed abundances.
Furthermore, we find that Einstein radii defined via the area enclosed
by the critical curve are $10-30\%$ larger than those defined via radial
surface mass density profiles. The contributions of radial and
tangential arcs to the radial distribution of arcs can be clearly
distinguished. The radial distribution of tangential arcs is very broad
and extends out to several Einstein radii. Thus, individual arcs are
not well suited for constraining Einstein radii.
\end{abstract}

\begin{keywords}
gravitational lensing -- dark matter -- large-scale structure of the
Universe -- galaxies: clusters: general -- cosmology: theory -- methods: numerical
\end{keywords}

\section{Introduction}
\label{sec:Introduction}

In studies of strong gravitational lensing, the mass distribution
deflecting the light is typically assumed to be concentrated in a single
region that is much smaller than the distance between source and
observer. The extension of the matter distribution along the
line-of-sight can be neglected in this case. The light deflection may
then be described by projecting the lens' mass distribution onto a
single plane and by assuming that light is deflected only when it passes
this lens-plane and propagates along straight lines otherwise. In this
so-called thin-lens approximation, the deflection angle is only
sensitive to the projected surface mass distribution of the lens.

This approximation has been widely used to study many aspects of
gravitational lensing. In particular, it was adopted to investigate the
efficiency of galaxy clusters to form giant arcs. However, studies
employing it together with spherically symmetric analytic cluster models
\citep[e.g.][]{WuMao1996,HattoriWatanabeYamashita1997,MolikawaEtal1999}
found significantly lower numbers of giant arcs than observed. The
discrepancy is reduced when considering triaxially shaped halos
\citep{OguriLeeSuto2003,MeneghettiEtal2007}. More realistic lens
models that also contain substructure can be obtained from
cosmological simulations. They were used by several authors to
investigate the strong-lensing efficiency of clusters
\citep[see e.g.][]{BartelmannWeiss1994}. One of the findings was that
even when using such detailed numerical models, the number of observed
giant arcs is significantly larger than the value expected in a
$\Lambda$CDM cosmology \citep{BartelmannEtal1998}. The discrepancy can
be reduced by taking cluster mergers \citep{TorriEtal2004}, baryonic
physics \citep{PuchweinEtal2005,WambsganssOstrikerBode2008,RozoEtal2008}
and a broad source redshift distribution
\citep{WambsganssBodeOstriker2004,LiEtal2005} into account. However, the
question whether or not the number of observed giant arcs is compatible
with a $\Lambda$CDM cosmology is still not settled, especially for a
low value of the matter power spectrum normalisation $\sigma_8$ \citep{FedeliEtal2008} as suggested by recent
cosmological constraints \citep{KomatsuEtal2009}.

Almost all of the studies mentioned above employed the thin-lens
approximation and considered galaxy clusters as isolated gravitational
lenses. In reality, however, clusters are embedded in a cosmic web of
large-scale structures. In addition, other collapsed objects like
galaxies or galaxy groups may be between a lensed background galaxy and
the cluster or between the cluster and the observer and thus contribute
to the lensing magnification and distortion. The effect of such
additional structures along the line-of-sight has been ignored in most
cluster-lensing studies.

A notable exception, in which the impact of such secondary structures
was specifically investigated, is \cite{WambsganssBodeOstriker2005}.
Using a large cosmological simulation and multiple lens planes between
source and observer, it was found that for high source redshift a
significant fraction of the lines-of-sight with a convergence exceeding
the critical value of unity for strong lensing/multiple image systems,
does so only due to a contribution from a second subdominant lens-plane.
This suggests that additional structures along the line-of-sight can,
indeed, be important for strong lensing of high-redshift sources. From
these results, it is, however, difficult to quantify the impact of such
secondary structures on the abundance of giant arcs.

On the other hand, \cite{HennawiEtal2007} also used multiple lens planes
in their analysis, but found that for sources at redshift $z_{\rm s}=2$,
additional structures along the line-of-sight only mildly ($\leq 7\%$)
affect the number of multiple-image systems with image separations
larger than 10 arcsec. But then, the largest effect was found for even
higher source redshift in \cite{WambsganssBodeOstriker2005} and there
may be a significant contribution from smaller-separation strong-lensing
systems.

There were also several studies investigating the impact of cluster
galaxies on the lensing properties of clusters.
\cite{MeneghettiEtal2000} and \cite{FloresMallerPrimack2000} found that
the lensing efficiency of clusters is only mildly affected by its member
galaxies. Brightest cluster galaxies, however, were not specifically
accounted for in these studies, but were shown to be able to raise
lensing
cross-sections by $\sim50\%$ in
\cite{MeneghettiBartelmannMoscardini2003_cD_gals}.
\cite{DalalHolderHennawi2004} showed that while the effect of galaxies
on wide-separation giant arcs is small, it becomes significant for
smaller $< 15''$ separation arcs.
\cite{HilbertEtal2008_StrongLensing_II} found that for point-like
circular sources, galaxies can boost the number of images with
length-to-width ratios exceeding 10 by up to a factor of 2 even when
considering only images with separations $> 5''$. 

In this work we investigate in detail how strong cluster lensing is
affected by secondary matter structures along the line-of-sight and by
both cluster and interloping galaxies. We carefully quantify their
impact and explore its dependence on cluster mass, cluster redshift, and
source redshift. We also study how the total lensing efficiency of a
realistic cluster population is affected. Our strong-lensing simulations
employ ray-tracing along a backward light cone constructed from the
Millennium simulation and use elliptical sources with sizes matching
observations. The impact of galaxies is studied using a semi-analytic
catalog that provides realistic galaxy properties and positions.

We also compare different definitions of the Einstein radius, as it is a
practical quantity for characterizing the strength of a gravitational
lens. Recently, \cite{OguriBlandford2009} suggested to use the
statistics of large Einstein radii as a cosmological probe, e.g. for
measuring $\sigma_{\rm 8}$. Here, we present the Einstein radius
distribution of the Millennium simulation clusters which can be used to
calibrate such methods. Finally, we investigate at what cluster-centric
radii giant arcs are most likely to be found.

The ray-tracing and strong-lensing simulation codes are introduced in
Sect. \ref{sec:methods}. Our results are presented in Sect.
\ref{sec:results} and summarized in Sect. \ref{sec:summary}.

\section{Methods}
\label{sec:methods}

\subsection{Ray-tracing}
\label{sec:ray_tracing}

We employ ray-tracing through the Millennium Simulation
\citep{SpringelEtal2005_Millennium}, a large $N$-body simulation of
cosmic structure formation, to study the light deflection by matter
structures in the observer's backward light cone. The Millennium
Simulation assumes a flat $\Lambda$CDM cosmology with a matter density
$\Omega_\mrm{M}=0.25$ (in units of the critical density), a cosmological
constant with energy density $\Omega_\Lambda=0.75$, a Hubble constant of
$h=0.73$ (in units of $100\,\km\,\mrm{s}^{-1}\Mpc^{-1}$), and a
scale-invariant initial density power spectrum with normalisation
$\sigma_8=0.9$. The simulation used a TreePM algorithm
\citep{Springel2005_GADGET2} with $10^{10}$ particles of mass
$m_\mathrm{p}=8.6 \times 10^8 h^{-1}\Msolar$ and a force softening of
$5h^{-1}\kpc$ comoving to simulate the structure formation in a cubic
region of $L=500 h^{-1}\,\mathrm{Mpc}$ comoving side length. Simulation
snapshots were stored on disk at 64 output times between redshift
$z=127$ and $z=0$.

We use the multiple-lens-plane algorithm described in
\citet{HilbertEtal2008_Raytracing} to calculate the gravitational
deflection of light by matter inhomogeneities between the source and the
observer. The observer's backward light cone is divided into redshift
slices. The matter content of each  slice is then projected onto a lens
plane transverse to the line-of-sight. Light rays are traced back from
the observer to their source under the assumption that photons propagate
unperturbed between these lens planes, but photons passing through a
plane are deflected by an amount that is determined by the projected
matter distribution on the plane.

The distribution of the dark matter in the observer's light cone is
generated directly from the particle data of the Millennium Simulation.
We use one redshift slice for each simulation snapshot. Each slice is
filled with the simulation particles of the corresponding snapshot. The
particles are projected onto a hierarchy of meshes on the corresponding
lens plane. The matter distribution on the meshes is smoothed by an
adaptive scheme that retains a resolution of about $5h^{-1}\,\kpc$
comoving in dense regions (e.g. in the centres of massive dark-matter
halos). The light deflection due to the dark matter is then calculated
from the smoothed distribution with Fast-Fourier-Transform methods,
finite differencing, and bi-linear interpolation.

The distribution of the luminous matter in the light cone is inferred
from the semi-analytic galaxy-formation model by
\citet{DeLuciaBlaizot2007}. The galaxies within each redshift slice are
projected onto the same lens plane as the simulation particles. For each
galaxy, the projected stellar matter is assumed to follow a
de-Vaucouleurs profile~\citep{DeVaucouleurs1948} for the bulge
component, and an exponential surface density profile for the disc
component. The light deflections induced by stars in the galaxies are
then calculated by analytic expressions.\footnote{Details about the
treatment of the light deflection by the stellar matter can be found in
\citet{HilbertEtal2008_StrongLensing_II}.}

In order to study the lensing properties of galaxy groups and clusters,
we select a sample of massive dark-matter halos at three lens redshifts
$\zL=0.28$, 0.62, and 0.99 in the Millennium Simulation. For each
halo, we set up light rays starting from the observer with directions on
a regular grid of $1024\times1024$ rays in a $6\arcmin\times6\arcmin$
field-of-view. The observer position is chosen such that the halo centre
is contained in the light cone spanned by the rays. The rays are traced
back through the series of lens planes, and the ray positions on each
lens plane, which also serve as source planes, are recorded. These ray
positions are then used to create lensed galaxy images as described in
Sect.~\ref{sec:strong_lensing_simulations}.

In this work, we want to investigate the influence of the stellar mass
in galaxies on the strong-lensing properties of groups and clusters. We
thus 
perform the ray-tracing not only as described above, but also by
ignoring the contribution from the stellar matter in galaxies to the
light deflection. We then compare the results we obtain by either
including or ignoring light deflection by the luminous matter.

Furthermore we want to quantify the contribution from additional matter
structures along the line-of-sight. Therefore in addition to the
complete multiple-lens-plane ray-tracing (hereafter \textit{full
ray-tracing}), we also perform ray-tracing in the following ways: In one
set of ray-tracing simulations, we only take into account the light
deflection by the lens plane containing the selected dark-matter halo
(\textit{single-plane ray-tracing}). These ray-tracing simulations
still include the effects of correlated structures close to the
lensing cluster but neglect contributions from matter structures at
different redshifts. In addition we perform another set of simulations,
in which only the matter in a cubic region of $4h^{-1}\,\Mpc$ comoving
side length around the halo centre is taken into account in the
ray-tracing (\textit{cluster-only ray-tracing}).

\subsection{Strong-lensing simulations}
\label{sec:strong_lensing_simulations}

We investigate strong lensing in these simulated light cones by placing
sources on six different source planes at redshifts $\zS=1.0$, 1.5,
2.1, 3.1, 4.2, and 5.7. We then find the images of each source using the
ray positions from the ray-tracing described above and calculate the
length $L$, width $W$ and magnification for each image. This allows a
statistical characterisation of the strong-lensing properties of the
matter structures in each light cone.

For finding the images of a large number of source galaxies, we follow
the method introduced by \cite{Miralda-Escude1993a,Miralda-Escude1993b}
and adapted to non-analytic models by \cite{BartelmannWeiss1994} and
\cite{BartelmannEtal1995}. The algorithm is described in some detail in
\cite{PuchweinEtal2005}. A previous version is also discussed in
\cite{MeneghettiEtal2000}.

First, the critical curves and caustics for each lens are determined by
checking where the determinant of the Jacobian of the lens mapping
changes its sign. Then, the algorithm adaptively places source galaxies
on the source plane such that regions near caustics are sampled with a
higher source density than regions far from caustics. A statistical
weight for each source accounts for the varying source density.
Source galaxies are modelled as ellipses with equivalent radii of 0.85,
0.65, 0.45, 0.3, 0.25, and $0.25\,\arcsect$ at redshifts $\zS=1.0$, 1.5,
2.1, 3.1, 4.2, and 5.7 respectively.\footnote{These source radii agree
well with measured half-light radii of galaxies at these redshifts
\citep{Ferguson2004}.}
Source ellipticities are assigned by randomly drawing ratios of major to
minor axis from the interval [1,2].

Each of the ray-traced fields-of-view is covered by a grid of $4096
\times 4096$ pixels in the image plane. The source-plane positions of
these pixels are calculated by bi-linear interpolation between the
source-plane positions of the $1024\times1024$ rays that were traced
through the simulation as described in Sect.~\ref{sec:ray_tracing}. All
image pixels of a source are then found by checking which of the grid
pixels correspond to a source-plane position that is enclosed by the
ellipse representing the considered source galaxy. The image pixels are
then grouped into images.

The properties of an image are obtained by finding the image pixel $\pixel_1$
that falls closest to the source centre when mapped to the source plane,
the pixel $\pixel_2$ that is the farthest from $\pixel_1$, and the pixel $\pixel_3$ that is
the farthest from $\pixel_2$. We then fit a circle through these three points
and use the arc length from $\pixel_2$ to $\pixel_3$ as the length of the image.
The image area is calculated directly from the number of image pixels,
while its perimeter is obtained by walking along the boundary pixels
and summing up their mutual distances. We apply the same resolution
corrections to image length and perimeter as in \cite{PuchweinEtal2005},
although they are less important here due to the better image plane grid
resolution.\footnote{We also performed several strong lensing
simulations with an increased $8192\times8192$ grid resolution to
check that results are robust even for the smallest source sizes.
In all cases, strong lensing cross sections changed by less
than 10\% compared to the lower resolution results. Also note that any 
small residual resolution dependence will affect the results obtained
by different kinds of ray-tracing in the same way.} We then compute a
simple geometric figure (ellipse, circle,
rectangle or ring) with equal area and length to determine the image
width $W$, which is approximated by the minor axis of the ellipse, the
diameter of the circle, the smaller side of the rectangle or the width
of the ring, respectively. We choose the type of the figure by comparing
its circumference to the perimeter of the image.

Finally, for all light cones and each source redshift, we quantify the
efficiency to produce arcs, which we define as images with a
length-to-width ratio $L/W>7.5$. More precisely, we determine
strong-lensing cross-sections $\sigma_{L/W>7.5}$ by summing up the
statistical weights of the sources that have images with $L/W>7.5$ and
calculating the corresponding area in the source plane. If there is more
than one arc for a source, we multiply the statistical weight of this
source by the number of arcs.

\section{Results}
\label{sec:results}

In order to quantify the impact of additional matter structures along
the line-of-sight on strong cluster lensing, we selected in total 100
cluster-sized halos, only by their mass, from the Millennium
simulation. Table
\ref{tab:mass_bins} summarizes their distribution in redshift $\zL$ and halo
mass $\Mcluster$.\footnote{Throughout this paper, we define cluster
mass as the mass within a spherical region with a mean density 200
times the critical density of the universe at the cluster redshift.}
For each cluster, we then performed \textit{full},
\textit{single-plane}, and
\textit{cluster-only} ray-tracing simulations, as described in
Sect.~\ref{sec:ray_tracing} and calculated the corresponding
strong-lensing cross-sections as detailed in
Sect.~\ref{sec:strong_lensing_simulations}.
Comparing them to each other allows us to clearly pin down the effect of
additional line-of-sight structures on cluster lensing
properties and giant arc statistics.

For all selected clusters, we also performed
\textit{full} and \textit{cluster-only} ray-tracing simulations that
ignored the light deflection by the stellar matter in both cluster and
interloping galaxies. This enables us to also quantify the impact of the
stellar component of galaxies on strong cluster lensing.

\begin{table}
\begin{tabular}{lccccc}
\hline
mass bin & (a) & (b) & (c) & (d) & (e) \\
$\Mcluster$ [$10^{14} h^{-1}\,\Msolar$]& 1-2 & 2-4 & 4-8 & 8-16 & $>16$ \\
\hline
\#clusters in bin & ~ & ~ & ~ & ~ & ~ \\
at $\zL$=0.28 & 10 & 10 & 10 & 8 & 1 \\
at $\zL$=0.62 & 10 & 10 & 10 & 4 & 0 \\
at $\zL$=0.99 & 10 & 10 & 7  & 0 & 0 \\
\hline
\end{tabular}
\caption{Mass range and number of studied clusters within the mass bins
at redshifts $\zL=0.28$, 0.62, and 0.99. We aim to use 10 clusters per
bin, but there are $<10$ clusters in the Millennium simulation in the
mass range of some bins. Also within each bin, clusters are roughly
constantly spaced in logarithmic cluster mass.}
\label{tab:mass_bins}
\end{table}

\subsection{The impact of additional structures along the line-of-sight}
\label{sec:additional_structures}

\begin{figure*}
\centerline{\includegraphics[width=0.95\linewidth]{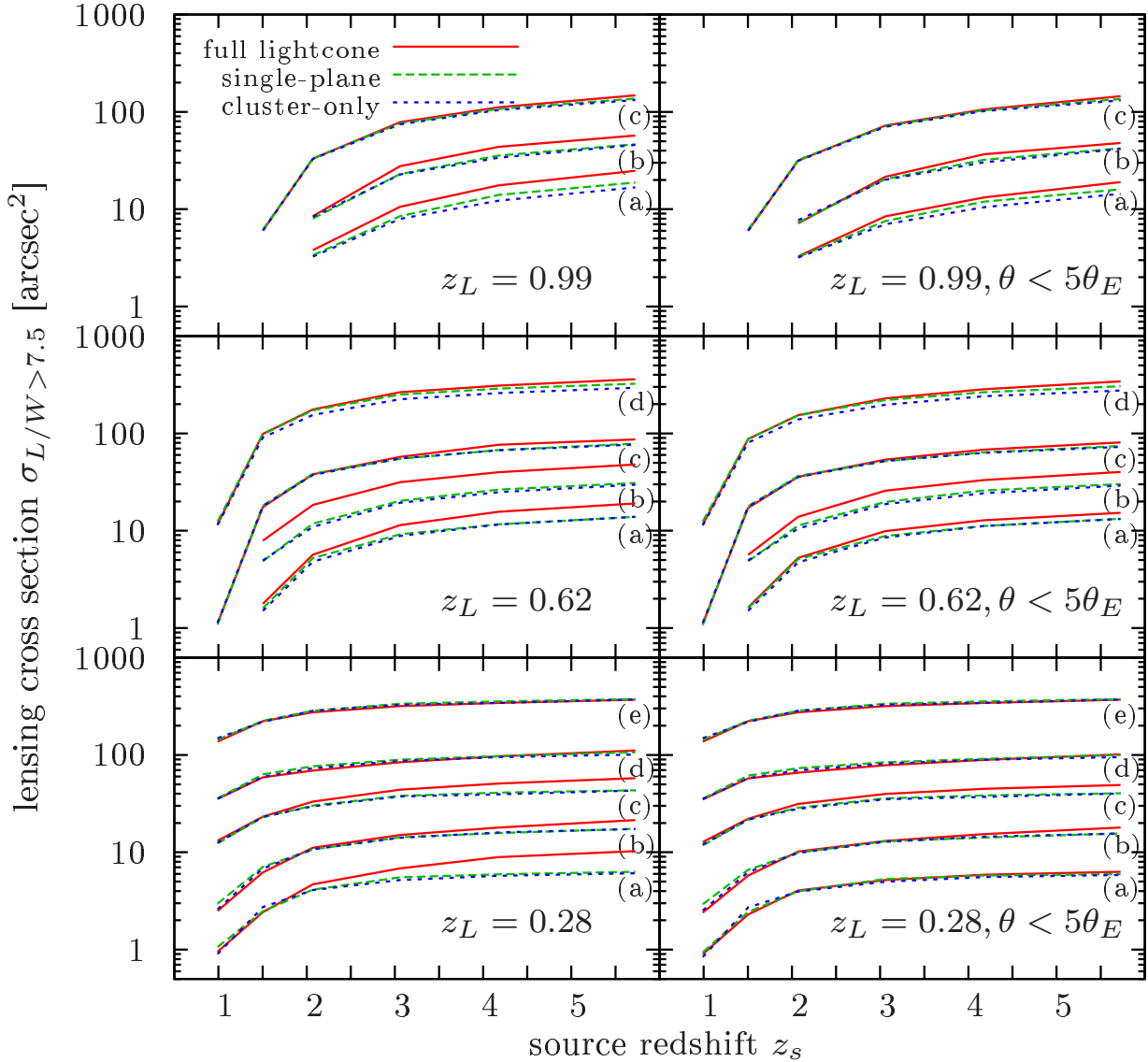}}
\caption{Average strong-lensing cross-sections $\sigma_{L/W>7.5}$ of
clusters in mass bins (a)-(e) (see Table
\ref{tab:mass_bins}). The cross-sections were obtained using full
light cones (\textit{red solid lines}), only one lens plane containing
the selected cluster (\textit{green long-dashed}), and light deflection
only by the selected cluster (\textit{blue short-dashed}). Results for
clusters at redshift $\zL$=0.28 (\textit{bottom panels}), $\zL=0.62$ (\textit{middle panels}), and $\zL=0.99$ (\textit{top panels}) are shown. In the \textit{left panels}, arcs in the whole $6\arcmin \times 6 \arcmin$ field were considered, while in the \textit{right panels}, only arcs within 5 Einstein radii were included.}
\label{fig:all_massbins}
\end{figure*}

In Fig.~\ref{fig:all_massbins}, we show the average strong-lensing
cross-sections $\sigma_{L/W>7.5}$ for all cluster mass bins given in
Table \ref{tab:mass_bins} and all considered source and lens redshifts.
For each mass bin and lens redshift, a set of three curves illustrates
the results obtained when using the \emph{full},
\emph{single-plane}, and \emph{cluster-only} ray-tracing. In all
cases, the stellar matter in galaxies was accounted for in the
ray-tracing. In the left panels, all suitable arcs in the whole
$6\arcmin\times6\arcmin$ ray-tracing fields where considered
for calculating $\sigma_{L/W>7.5}$. 

Probably the most obvious result is that the average cross-section is a
strongly increasing function of source redshift for all mass bins, lens
redshifts, and types of ray-tracing. Clusters are, thus, more
efficient in lensing source galaxies at higher redshift. This finding is
in good agreement with previous studies \citep{WambsganssBodeOstriker2004,LiEtal2005,HilbertEtal2007_StrongLensing}. 

A more careful inspection of the results reveals that for all considered
lens redshifts, the cross-sections of the most massive clusters (uppermost
set of curves in each panel) are only very mildly affected by additional
matter structures along the line-of-sight. The results obtained using
\textit{full} ray-tracing are very similar to those found by
\textit{single-plane} and \textit{cluster-only} ray-tracing for such
massive clusters.

The effects of additional matter structures along the line-of-sight are more important for less massive clusters and for sources at high redshift.
There, the cross-sections obtained when using the complete multiple-lens
plane
ray-tracing are on average $\sim 20-40\%$ larger than those found by
considering only light deflection by the cluster. On the other hand, the
difference between results from \textit{single-plane} and
\textit{cluster-only} ray-tracing are still very small. Thus the
increase in lensing efficiency found when accounting for all the matter
in the light cone is not due to correlated structures near the cluster, but due to uncorrelated structures along the
line-of-sight.

So far, all suitable arcs, i.e. with a length-to-width ratio exceeding
7.5, found in the ray-traced field were counted for the computation of a
cluster's strong-lensing cross-section. However, an observer might not
assign all these arcs to the selected cluster, but also to other objects
that happen to be in the ray-traced field. Neglecting this is unlikely
to significantly bias results for the most massive clusters. However,
the smallest clusters in our sample are in some cases not even the most
prominent lens in the field-of-view. Thus by assigning all found arcs
to such a cluster, one would overpredict its lensing efficiency,
especially when using the \textit{full} ray-tracing. 

To prevent such misassignments from
biasing our results, we repeated the computation of $\sigma_{L/W>7.5}$
for all cluster, but this time counting only arcs within 5 Einstein
radii from the selected cluster's centre. Note that here, we define the
Einstein radius $\theta_E$ by
\begin{equation}
\theta_E = \sqrt{\frac A \pi},
\label{eq:def_einstein_radius_crit_curve}
\end{equation}
where $A$ is the solid angle enclosed by the cluster's outer critical
curve and calculated using \textit{cluster-only} ray-tracing. In this
way, we largely avoid assignments to a wrong object. In rare cases of ongoing mergers or very elliptic clusters with
substructure, we might, however, miss some arcs that are very distant from the cluster
centre but should probably be assigned to it. We employ this
cut in image separation troughout the remainder of this paper.

Results obtained in
this way are shown in the right panels of Fig. \ref{fig:all_massbins}.
As expected, the cross-sections of the most massive clusters are almost
unchanged. For the smallest clusters on the other hand, the lensing
cross-section obtained by \textit{full} ray-tracing is reduced and
becomes more similar to the \textit{cluster-only} results. Thus, when
counting only arcs within 5 Einstein radii, most small clusters
are not significantly affected by additional structures along the
line-of-sight. The effect is biggest for intermediate mass clusters,
where cross-sections increase on average by roughly $\sim 15-30\%$.

\begin{figure}
\centerline{\includegraphics[width=\linewidth]{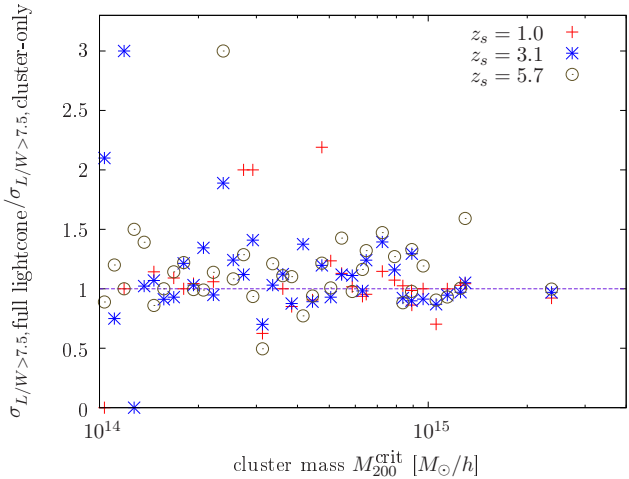}}
\caption{Ratio of the strong-lensing cross-section $\sigma_{L/W>7.5}$
found when using the \textit{full} multiple-lens-plane ray-tracing to
the value
obtained when considering only light deflection by the selected cluster.
Results are shown as a function of cluster mass for all $\zL$=0.28
clusters in our sample. Three different source redshifts, $\zS=1.0$, 3.1, and 5.7 were used. Only arcs within 5 Einstein radii were included in the analysis.}
\label{fig:cross_sec_ratio}
\end{figure}

For individual clusters, however, effects can be significantly larger.
We illustrate this in Fig.~\ref{fig:cross_sec_ratio}, which shows the
ratio of the cross-section obtained by the \textit{full} ray-tracing to
the values found by \textit{cluster-only} ray-tracing. For the sake of
clarity, only the results for the $\zL$=0.28 clusters and three
different source redshifts are shown. Again, only arcs within 5 Einstein
radii were counted. Boosts of the cross-sections by additional
matter structures of up to $50\%$ happen frequently. There are also a
couple of clusters with significantly larger ratios. For most of them,
however, absolute values of the cross-sections are quite small, so that
a small absolute increase results in a large relative boost.

A particularly impressive example is the second most massive
cluster shown in the figure. In spite of its enormous mass, secondary
structures along the line-of-sight boost its cross-section for
lensing high-redshift sources by roughly $60\%$. A large part of this 
increase is due to three group-sized halos, two at redshift 1.5 and one
at redshift 1.63, which are close to the cluster in projection. In
addition, there is a significant effect due to a $2\times 10^{12}
\Msolar/h$ halo that is also at redshift 1.63. Normally, one would not
expect such a small object to have a noticeable impact on the lensing
properties of a cluster whose mass is almost three orders of magnitude
larger. This halo is, however, itself strongly lensed by the foreground
galaxy cluster. The peak in the lensing convergence corresponding
to it is shaped like a long thin tangential arc. Due
to the lensing magnification of this halo, its light deflection affects
a much larger number of lines-of-sight. In this way, this small object
can significantly alter the critical curve of the cluster. Also see
\cite{HilbertEtal2008_Raytracing} for a discussion of how
multiple-lens-plane ray-tracing can be used to study lens-lens coupling
effects in weak lensing. Finally, there is also some contribution
from several halos with masses of a few $10^{12}\Msolar/h$ and
redshifts between $\zL=$1.7 and 3.3. Remarkably, the importance of
additional line-of-sight structures is already strongly reduced for
sources at a somewhat lower redshift of $z_{\rm s}$=3.1. This is
because the efficiencies of lenses at $\zL > 1.5$ significantly
increase with source redshift between $z_{\rm s}$=3.1 and 5.7, while
there is only a marginal gain for a lens at the much lower cluster
redshift of $\zL=$0.28.

There are also several clusters which have a smaller cross-section when
including the effects of additional structures along the line-of-sight.
However, the decrease is less than $10\%$ in most cases. Note
that such clusters are not necessarily projected onto underdense
regions, as we do not find a clear correlation between change in
cross-section and change in average convergence within 5 Einstein
radii.\footnote{The region inside which the average convergence was
computed was kept fixed, as we used the Einstein radius obtained from
\textit{cluster-only} ray-tracing in both cases.} In most cases where we
find significantly smaller cross-sections when including all structures
in the light cone, it is the `external' shear
caused by the secondary structures along the line-of-sight that
counteracts the cluster shear on potential arcs. In many cases, this
lowers their length-to-width ratios below the adopted threshold value of
7.5.

\begin{figure*}
\centerline{\includegraphics[width=0.95\linewidth]
{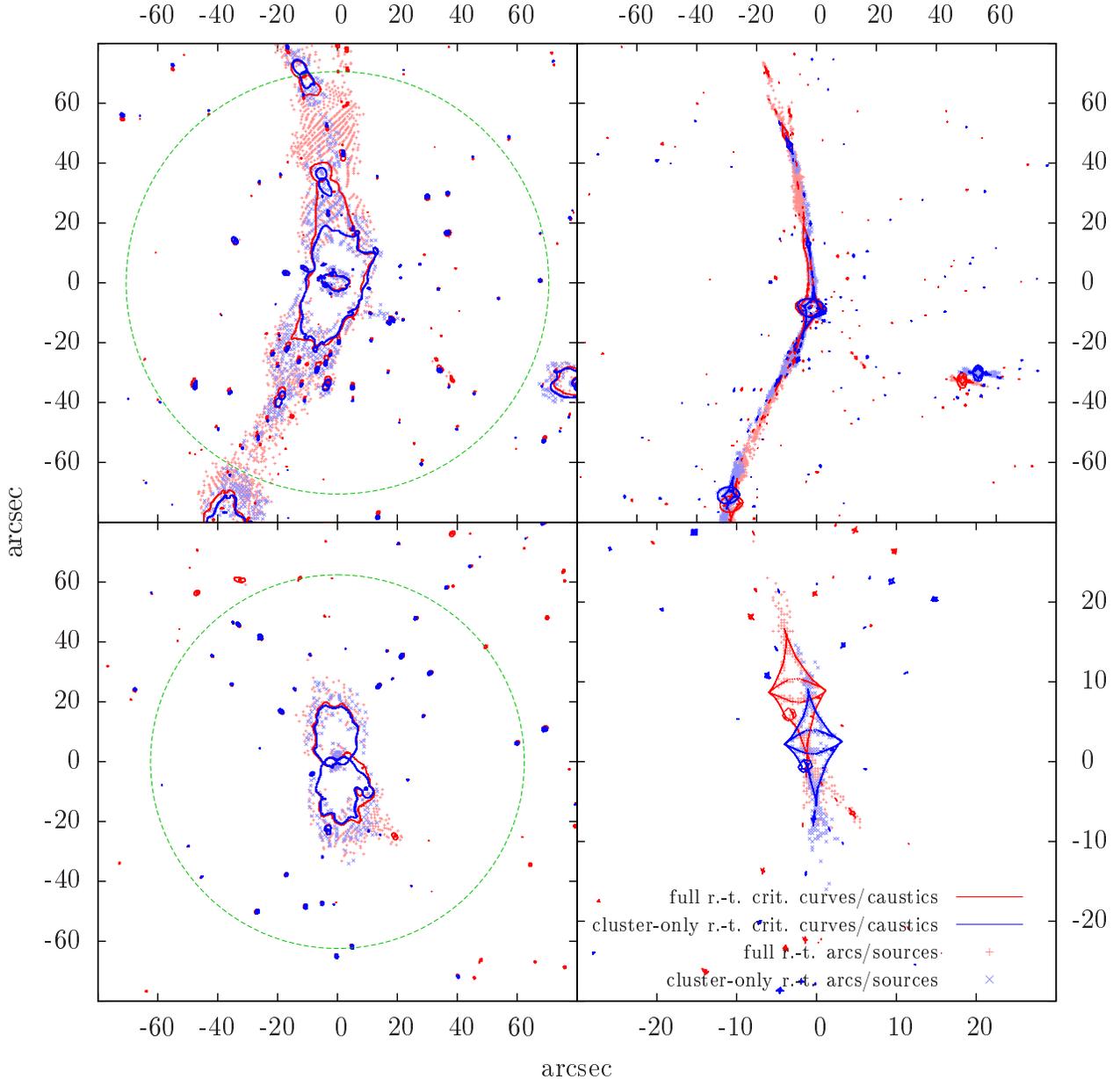}}
\caption{Critical curves (\textit{left panels}) and caustics
(\textit{right panels}) of a $10^{15} h^{-1}\,\Msolar$ cluster at $\zL=0.62$ (\textit{upper panels}) and a $7\times10^{14} h^{-1}\,\Msolar$
cluster at $\zL=0.28$ (\textit{lower panels}). Curves obtained by
considering light-deflection only by the cluster (\textit{blue}) as well
as by all matter structures along the line-of-sight (\textit{red}) are
shown. A source redshift of $\zS=3.1$ was assumed. The cluster
shown in the \textit{upper panels} is a prominent example for an object
whose lensing efficiency is strongly boosted by additional structures
along the line-of-sight. Its cross-section increases by almost a factor
of 2. The \textit{lower panels} show a more typical example, where the
cross-section increases by $\sim 30\%$. In both cases only giant arcs
within 5 Einstein radii (\textit{green circles}) were counted. In the
\textit{left panels}, the centres (defined by pixel $\pixel_1$ as
described in
Sect. \ref{sec:strong_lensing_simulations}) of all arcs with $L/W>7.5$
found in the strong-lensing simulation are shown. In the \textit{right
panels}, the positions of the sources that give rise to these arcs are
indicated. Note that for the sake of clarity, we choose a different scale
in the \textit{bottom, right panel}.}
\label{fig:crit_curves_caustics}
\end{figure*}

Figure \ref{fig:crit_curves_caustics} shows the critical curves and
caustics of two clusters and illustrates how they change when the effects of additional structures along the line-of-sight are included. The first
cluster, shown in the figure's upper panels, is an example of a strongly
affected object. Its lensing cross-section is boosted by a factor of two, and the shape of its critical curve changes significantly.
The latter is stretched and merges with the critical curve of a subhalo.
Also shown are the centres of all giant arcs formed in the cluster's
\textit{full} and \textit{cluster-only} strong-lensing simulations as
well as the sources that give rise to them. They illustrate how the
shape of the cluster's strong-lensing cross-section changes and where
additional arcs appear. The second cluster is a more typical example.
Its cross-section increases by about $\sim 30\%$, mostly due to an
additional structure below and to the right of the cluster. The shape of
its critical curve and caustic are only mildly affected. One can however
clearly see that the caustics are shifted with respect to each other due
to light deflection by structures between the cluster and the source
plane.

\begin{figure}
\centerline{\includegraphics[width=\linewidth]{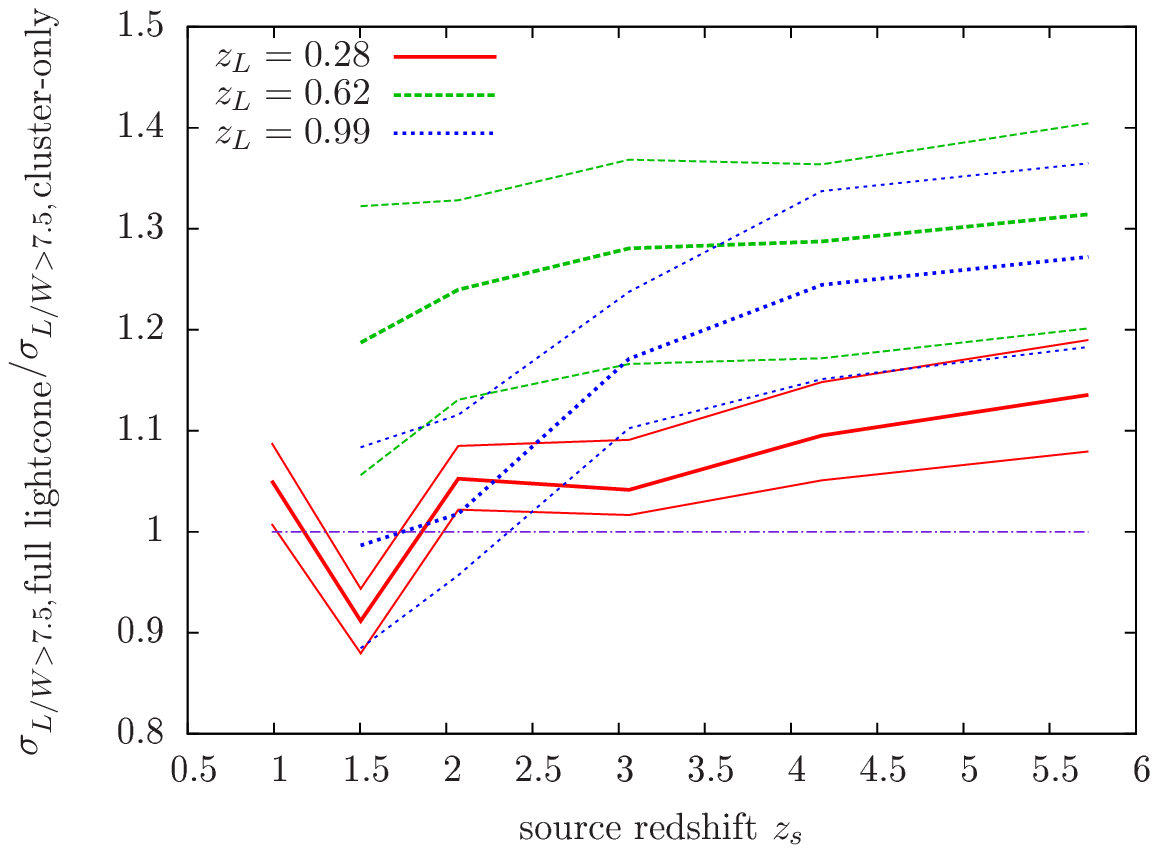}}
\caption{Impact of additional structures along the line-of-sight on the
combined strong-lensing cross-section of all Millennium simulation
clusters. Ratios of total cross-sections estimated from
\textit{full} and \textit{cluster-only} ray-tracing are shown as a
function of source redshift for the cluster populations at $\zL=0.28$,
0.62, and 0.99 (\textit{thick lines}). Only arcs within 5 Einstein radii
were included in the analysis. Also shown are $68\%$ confidence
intervals of the ratios (\textit{thin lines}) obtained by
bootstrapping techniques.}
\label{fig:all_cross_ratio}
\end{figure}

As mentioned above and illustrated in Fig. \ref{fig:all_massbins}, 
the impact of additional structures along the line-of-sight on cluster
strong-lensing cross-sections depends on cluster mass. Thus, one needs
to take the steep cluster mass function into account when investigating
how the total lensing efficiency of a realistic cluster population is
affected. However, instead of employing an analytic mass function for
that purpose, we directly use the full Millennium simulation halo
catalogue. We start from the cross-sections we calculated for our selected
cluster sample and then assign a cross-section to all other Millennium
simulation clusters, here defined as halos with
$\Mcluster>10^{14}h^{-1}\,\Msolar$, by interpolating linearly in
cluster mass. Summing over all clusters finally yields an estimate of
the total strong-lensing cross-section of the Millennium simulation
cluster population at $\zL=0.28$, 0.62, and 0.99. We calculated these
quantities based both on \textit{full} and on \textit{cluster-only}
ray-tracing. Figure \ref{fig:all_cross_ratio}
illustrates how additional structures along the line-of-sight affect the
lensing efficiency of the cluster population as a whole and thus the
contribution of clusters at these redshifts to the strong-lensing
optical depth. Clearly, the lensing efficiency for high-redshift sources
increases when taking such additional structures into account. This
effect is larger for intermediate and high redshift cluster populations,
where we find boosts up to $30\%$. For the $\zL=0.28$ clusters and high
redshift sources, the lensing efficiency increases by about $10\%$.

We also show $68\%$ confidence intervals of the total cross-section
ratios. For each cluster redshift, they were calculated from $10^4$
bootstrap cluster samples. Note that we bootstrap both the
clusters which we use for the cross-section
interpolation as well as the full cluster population in the
simulation. The former accounts for uncertainties due to the limited
number of objects for which we have performed strong-lensing simulations
and for noise in the cross-section calculation, while the latter
accounts for cosmic variance in the whole simulation.\footnote{Before
we applied bootstrapping, we
have performed a Monte-Carlo analysis that employed random
cross-sections drawn based on a rough estimate of the cross-section
cluster mass relation and its scatter to confirm that bootstrapping
gives reliable error estimates even when using such an interpolation
scheme to calculate total cross-section ratios.} Most uncertain is the
contribution from the large number of low mass clusters to the
strong-lensing optical depth. This is because their cross-sections are
small, especially for low source redshift, and thus the relative error
in their determination is larger.

Of course, the strong-lensing cross-section of individual
clusters does not only depend on cluster mass, but due to halo
triaxiality, also on orientation \citep{DalalHolderHennawi2004},
as well as on halo concentration \citep[see e.g.][]{HennawiEtal2007},
and dynamical state \citep{TorriEtal2004}. It is, therefore, not
suprising that the distribution of cross-sections is quiet broad even at
constant cluster mass. Obviously, the impact of structures along the
line-of-sight also depends on the line-of-sight's orientation and can
change significantly depending on it. Thus, while the curves given for
individual mass bins in Fig. \ref{fig:all_massbins} (and Fig.
\ref{fig:all_massbins_gals} in the next section) are very useful for
identifying trends with cluster mass, they may, due to the limited
number ($\le 10$) of objects per bin, not always be fully
representative for all cluster lenses in that mass range.
The results we give for the whole cluster population, on the other hand,
should be much more robust as a larger number of clusters is used for
their derivation. Also note that the bootstrap error estimates include
the effects mentioned above.

\subsection{The impact of the stellar mass in galaxies}
\label{sec:galaxies}

\begin{figure}
\centerline{\includegraphics[width=\linewidth]{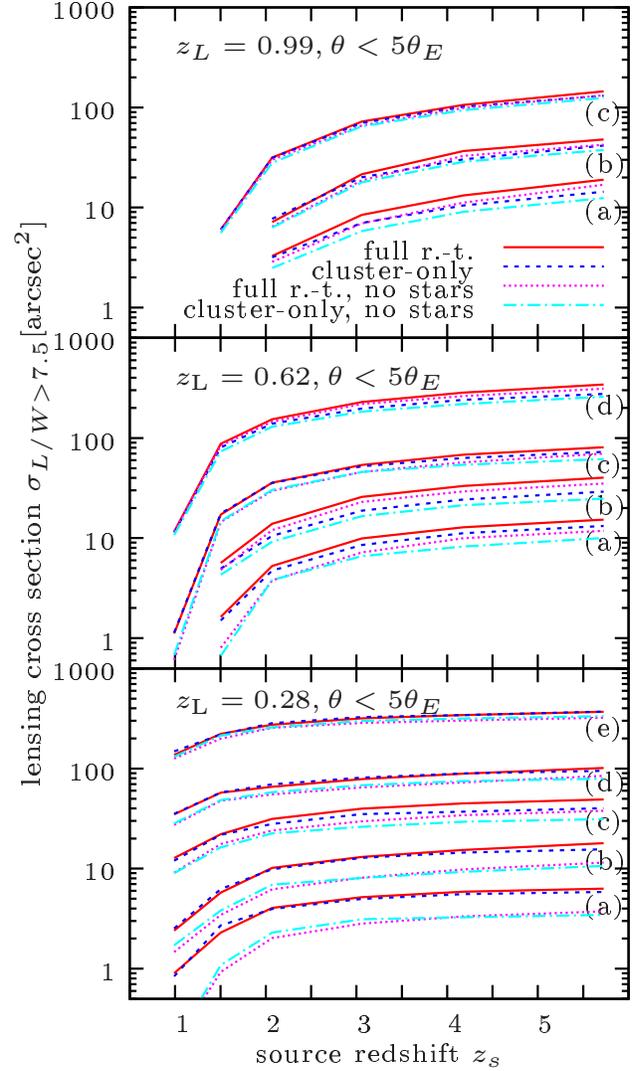}
}
\caption{Average strong-lensing cross-sections $\sigma_{L/W>7.5}$ of
clusters in mass bins (a)-(e) at cluster redshift
$\zL$=0.28 (\textit{bottom panel}), $\zL$=0.62
(\textit{middle panel}), and $\zL$=0.99 (\textit{top panel}). The
cross-sections were obtained using
\textit{full} ray-tracing and accounting for stellar matter (\textit{red
solid lines}), \textit{cluster-only} ray-tracing and accounting for
stellar matter (\textit{dark-blue, dashed}), \textit{full} ray-tracing
and ignoring stellar matter (\textit{magenta, dotted}),
\textit{cluster-only} ray-tracing and ignoring stellar matter
(\textit{light-blue,
dot-dashed}). Only arcs within 5 Einstein radii were counted. The
\textit{red solid} and \textit{dark-blue dashed} curves show the same
data as in the middle right and lower right panels of Fig.
\ref{fig:all_massbins}.}
\label{fig:all_massbins_gals}
\end{figure}

In Fig.~\ref{fig:all_massbins_gals}, we compare the strong-lensing
cross-sections calculated accounting for the light deflection by the
stellar component of galaxies to those found ignoring all stellar matter.
Average cross-sections for all the $\zL=0.28$, 0.62 and 0.99 cluster
mass bins listed in Table \ref{tab:mass_bins} are shown. Results
obtained by both \textit{full} and \textit{cluster-only} ray-tracing are
indicated. When accounting for stellar matter, the former include the
effects of all galaxies between source plane and observer while the
latter include only the effects of cluster galaxies. One can clearly see
that the stellar matter significantly increases the strong-lensing
cross-sections of clusters, especially for low-mass systems and for low
source redshift. In other words, the effect is largest for the most
inefficient lenses, while very massive clusters are only mildly
affected. It is also worth noting that the impact of stellar matter is
very similar when using \textit{full} and \textit{cluster-only}
ray-tracing. This means that most of the increase is due to the stellar
component of cluster galaxies, while interloping galaxies between source
and cluster or between cluster and observer play only a minor role.

\begin{figure}
\centerline{\includegraphics[width=\linewidth]{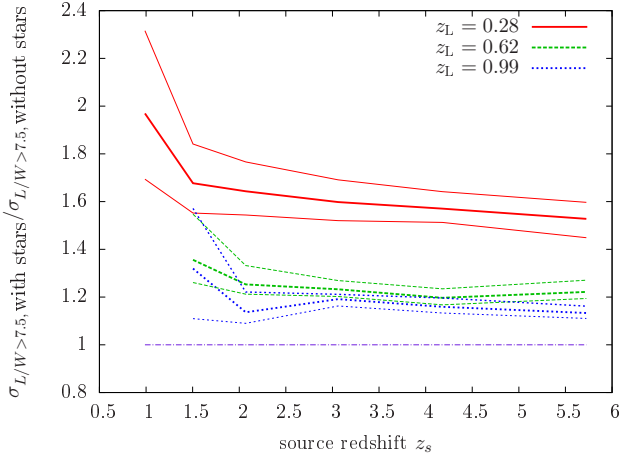}}
\caption{Impact of stellar matter on the combined strong-lensing
cross-section of all Millennium simulation clusters. Ratios of the total
cross-sections estimated accounting for and ignoring stellar matter are
shown as a function of source redshift for the cluster populations at
$\zL=0.28$, 0.62, and 0.99 (\textit{thick lines}). Also shown are $68\%$
confidence intervals of the ratios (\textit{thin lines}) obtained by
bootstrapping techniques. The analysis employed \textit{full}
ray-tracing. Only arcs within 5 Einstein radii were included.}
\label{fig:all_cross_ratio_gals}
\end{figure}

Also in this context, the impact on cluster lensing cross-sections
depends on cluster mass. We, thus, use the same method as in Sect.~\ref{sec:additional_structures} to estimate how the lensing efficiency
of a realistic cluster population is affected.
Figure~\ref{fig:all_cross_ratio_gals} illustrates the increase of the
combined
strong-lensing cross-section of all Millennium simulation clusters with
$\Mcluster>10^{14}h^{-1}\,\Msolar$. Relative changes due to the
stellar components of galaxies are shown as a function of source
redshift for all clusters at $\zL=0.28$, 0.62 and 0.99. We also
plotted the $68\%$ confidence intervals of the cross-section ratios
obtained by a bootstrap error analysis similar to that described in
Sect. \ref{sec:additional_structures}. \textit{Full} ray-tracing was
used to derive all the data shown in the figure. Stellar matter boosts
the lensing efficiency
of the $\zL=0.28$ cluster population by almost a factor of two for low
redshift sources. For high-redshift sources the strong-lensing
efficiency still increases by about $50\%$. The clusters at higher
redshift show the same trend in source redshift with somewhat larger
boosts for low redshift sources. However, the magnitude of the effect is
only about half of that found for the $\zL=0.28$ clusters.

In order to understand this difference better, we calculated the Einstein
radius of the average strong-lensing cluster for all lens redshifts.
More, precisely we assigned an Einstein radius to all Millennium
simulation clusters using the same interpolation method as for the
cross-sections. We then derived the cross-section-weighted Einstein
radii for the three cluster populations and all source redshifts. We
found that for sources at redshift 1.5 the Einstein radius of the
average lensing cluster is very similar for clusters at $\zL=0.28$ and
0.62, namely about $\sim 8$ arcsec, and larger in the latter case for
higher redshift sources. For the $\zL=0.99$ clusters it is compareable
or larger than for the $\zL=0.28$ clusters above a source redshift of
2.1. Thus, in spite of the much larger distance between lens and
observer the Einstein angle of strong-lensing selected clusters does
not decrease at high lens redshift. It corresponds, however, to a
larger physical distance at the cluster position. The stellar mass
within the critical curve is typically dominated by the central
galaxy, which is more concentrated than the dark matter halo. Thus, when
the physical size of the critical curve increases lensing becomes more
dark matter dominated and less affected by stellar matter.

Our results for the increase of the lensing cross
sections due to cluster galaxies are in good agreement with the results of \cite{HilbertEtal2008_StrongLensing_II} for point-like sources.
At first glance, these results may seem to be at odds with the
findings of \cite{MeneghettiEtal2000}. However, we find the largest
effect in small clusters which were not considered there. In addition,
we also accounted for brightest cluster galaxies which were also not
included in that study, but shown to be able to raise lensing
cross-sections by $\sim50\%$
in \cite{MeneghettiBartelmannMoscardini2003_cD_gals}.

\subsection{Optical depths for giant arcs}
\label{sec:optical_depths}

We can use our cluster sample to estimate the optical depth
$\tau_{L/W>7.5}$ for arcs with $L/W>7.5$ as a function of source
redshift $\zS$. For the considered cluster lens redshifts $\zL=0.28$,
0.62, and 0.99, we calculate the cross-sections
$\sigma_{L/W>7.5,i}$ of every halo in the simulation with mass
$\Mcluster\geq 10^{14}h^{-1}\,\Msolar$ by linearly interpolating the
cross-section as a function of logarithmic cluster mass between the
values of the ray-traced clusters.

We then compute an estimate of the
differential optical depth $\partial\tau_{L/W>7.5}/\partial\chiL$ at the
comoving line-of-sight distances $\chiL=\chi(\zL)$ of these three
redshifts $\zL$ by
\begin{equation}
  \frac{\partial\tau_{L/W>7.5}}{\partial\chiL} =
 \chiL^2\frac{\sum_{i}\sigma_{L/W>7.5,i}}{V_{\rm{box}}},
\end{equation}
where the $\sigma_{L/W>7.5,i}$ are assumed to be in
steradians and the summation goes over all clusters with $\Mcluster\geq
10^{14}h^{-1}\,\Msolar$ found in the
simulation output at redshift $\zL$. \footnote{Note that
the contribution of
low-mass clusters to the optical depth is more important for high
redshift sources. Our data suggest that for halos with
$\Mcluster<10^{14}h^{-1}\,\Msolar$, it becomes non-negligible for
$\zS\geq2.1$. This contribution is not included in this analysis.
On the other hand, it is likely to strongly depend on the exact
selection criteria for arcs. For example requiring some minimum value
for the image separation would largely suppress it.} $V_{\rm{box}}
= (500 h^{-1}\,\mathrm{Mpc})^3$ is the comoving volume of
the simulation box.

\begin{figure}
\centerline{\includegraphics[width=\linewidth]{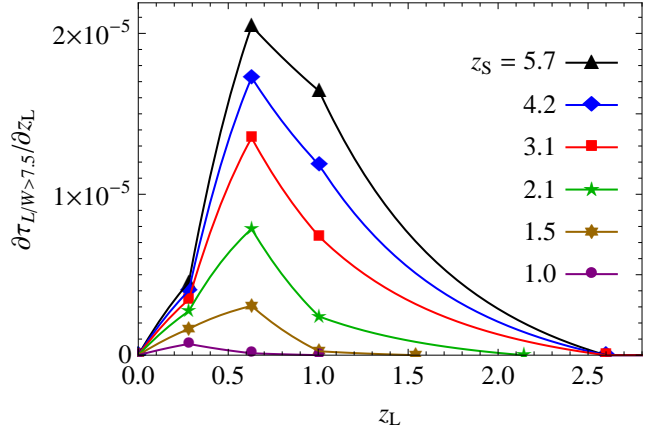}}
\caption{
\label{fig:d_tau_d_z}
Differential optical depths $\partial\tau_{L/W>7.5}/\partial\zL$
for arcs with length-to-width ratio $L/W>7.5$ as a function of lens
reshift $\zL$ for various source redshifts $\zS$. The
values obtained from \emph{full} ray-tracing including the stellar
matter and considering only arcs within 5 Einstein radii of the lens
cluster are shown as symbols. The lines indicate the differential
optical depths obtained from linear interpolation in the comoving
line-of-sight distance $\chiL$.
}
\end{figure}

Furthermore, we assume that $\partial\tau_{L/W>7.5}/\partial\chiL$
vanishes for $\zL=0$, $\zL\geq \zS$, and $\zL>2.6$ (there are no
clusters with $\Mcluster\geq 10^{14}h^{-1}\,\Msolar$ for $\zL>2.6$).
The contribution from other redshifts is then calculated by linear
interpolation in $\chiL$. The resulting differential optical depths
are illustrated for all considered source redshifts in
Fig.~\ref{fig:d_tau_d_z}. We show them as a function of the lens
redshift $\zL$ to make them more easily comparable to other works.

\begin{figure}
\centerline{\includegraphics[width=\linewidth]
{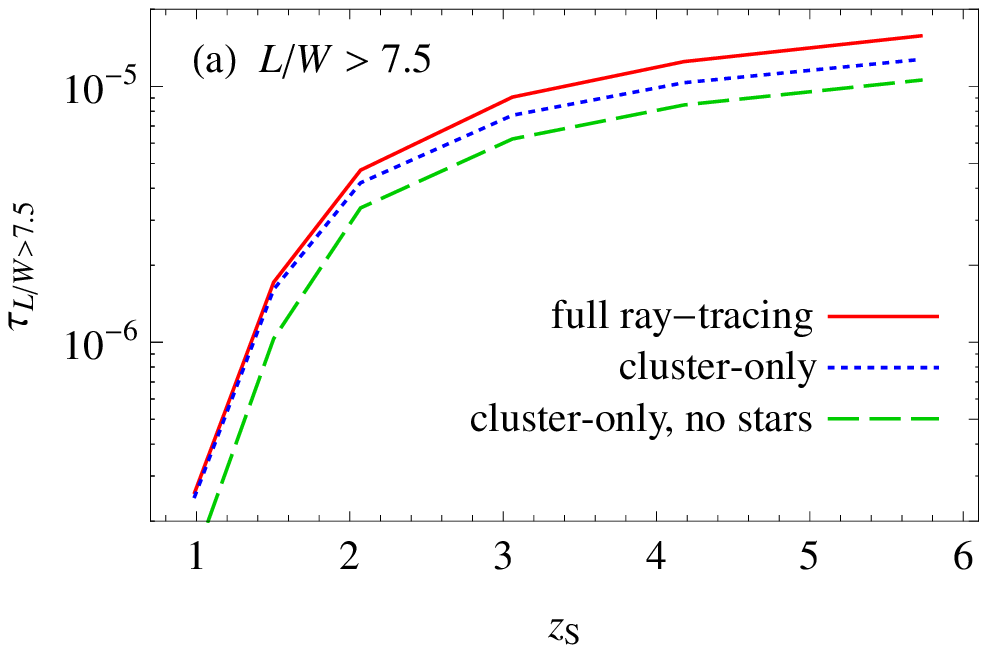}}
\centerline{\includegraphics[width=\linewidth]
{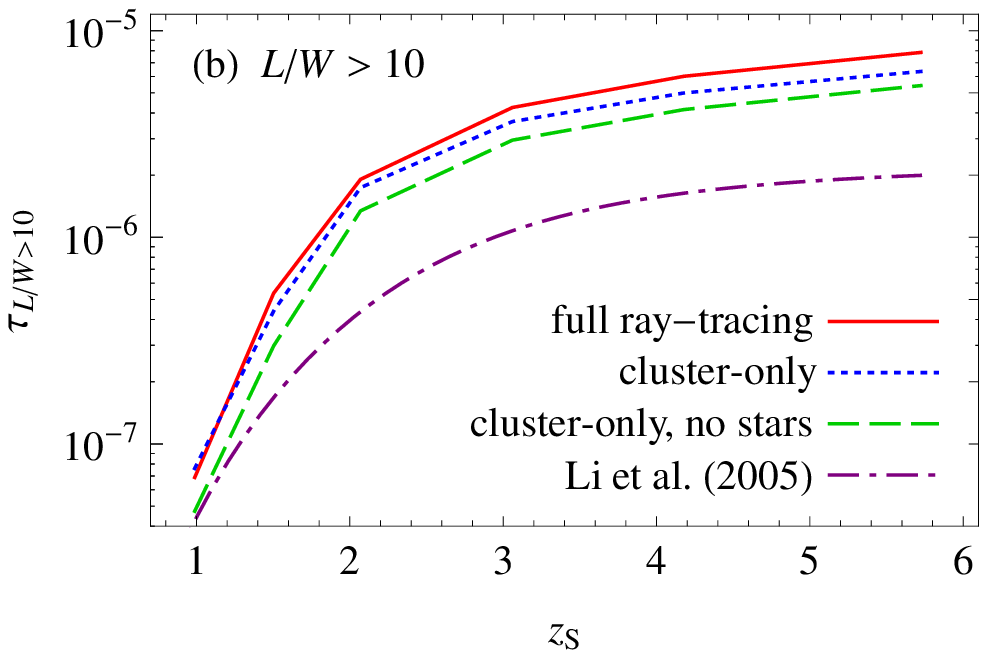}}
\caption{
\label{fig:tau}
Optical depths $\tau_{L/W>7.5}$ ((a),\textit{top panel}) and
$\tau_{L/W>10}$ ((b),\textit{bottom panel}), for arcs
with length-to-width ratio $L/W>7.5$ and $L/W>10$ as a function of
source redshift $\zS$. Shown are the values for \emph{full} ray-tracing
including the stellar matter (\textit{red solid lines}), for
\emph{cluster-only} ray-tracing with stars (\textit{blue dotted lines}),
and for \emph{cluster-only} ray-tracing without stars (\textit{green
dashed lines}). Only arcs within 5 Einstein radii were included. Also
shown is $\tau_{L/W>10}$ computed by \citet{LiEtal2005} for sources with
$1\arcsec$ effective diameter (\textit{purple dash-dotted line}).
}
\end{figure}

The optical depth $\tau_{L/W>7.5}$ is obtained by integrating
$\partial\tau_{L/W>7.5}/\partial\chiL$ between $\chiL=0$ and
$\chiL=\chi(\zS)$.
Similarly, we calculate the optical depth $\tau_{L/W>10}$ for arcs with
$L/W>10$. The optical depths resulting from \emph{full} ray-tracing
including the stellar matter in galaxies is compared to the estimates
based on \emph{cluster-only} ray-tracing with and without stellar matter
in Fig.~\ref{fig:tau}.

The qualitative behaviour of $\tau_{L/W>7.5}$ and $\tau_{L/W>10}$ is
very similar. The additional matter along the line-of-sight increases
the optical depths by $10\%$ for $\zS=2.1$ and by $25\%$ for $\zS=5.7$.
Hence, projection effects are more important for higher source
redshifts. The stars in galaxies enhance the optical depths by a factor
two for $\zS=1$ and cause an increase by $25\%$ for sources
at $\zS=5.7$. Hence, the stellar matter is more important for lower
source redshifts. Together, the stars and the additional matter along
the line-of-sight increase the optical depth by a factor 1.5 to 2.

For comparison with \citet{FedeliEtal2008}, we compute average optical
depths
\begin{equation}
	\bar{\tau}_{L/W>7.5}=\int\diff{\zS}	p_\mrm{S}(\zS)
\tau_{L/W>7.5}(\zS)
\end{equation}
using the source redshift distribution \citep[][]{SmailEtal1995}
\begin{equation}
	p_\mrm{S}(\zS)=\frac{3}{2}\zS^2\exp\left(-\zS^{3/2}\right).
\end{equation}
The resulting values $\bar{\tau}_{L/W>7.5}=2.5\times10^{-6}$ for
\textit{full}
ray-tracing and $\bar{\tau}_{L/W>7.5}=1.7\times10^{-6}$ for
\textit{cluster-only}
ray-tracing without stars are 4 to 6 times smaller than the optical
depths of \cite{FedeliEtal2008}. Part of the discrepancy may be due to
the fact that we assumed larger source radii at low redshifts.
In addition, there may be some difference in the lensing
properties between the simulated cluster population used in this work
and the analytic cluster population model employed there. In
particular, cluster mergers are accounted for in different ways. In our
simulations, cluster mergers naturally occur. However, we might be
limited by small-number statistics. \citet{FedeliEtal2008}, on
the other hand, explicitly include cluster mergers, assuming
that they proceed at the gravitational free-fall timescale and can 
be represented by two NFW halos approaching at constant velocity.
Without the merger contribution their optical depths would be $\sim 3$
times smaller and agree much better with our
results. However, \cite{HennawiEtal2007} question that mergers can
significantly boost optical depths for giant arcs. Considering all these
points, it seems likely that at least part of the discrepancy between
our results and those of \cite{FedeliEtal2008} is related to the way
cluster mergers are accounted for, i.~e. the importance of merger
boosting of strong lensing cross-sections may be overestimated in
\cite{FedeliEtal2008} and/or our selected cluster sample may still be
too small to fully account for strong boosts during rare major cluster
mergers.

In Fig.~\ref{fig:tau}(b), we compare our optical depths for arcs with
$L/W>10$ to those obtained by \citet{LiEtal2005}. The optical depths are
very similar for sources at $\zS=1$. However, our optical depths are 3
to 4
times larger for $\zS\geq2.1$. The fact that we used smaller source
radii at higher redshifts accounts only for an increase of
about $50\%$ \citep{LiEtal2005}. A reason for the remaining discrepancy
could be the significantly better spatial resolution of the
Millennium Simulation compared to the simulations used by
\citet{LiEtal2005}, which is likely to boost the lensing efficiencies
of poor clusters, whose contribution to the optical depth is most
important at high source redshift. Another reason may be the differences
in the assumed cosmology.

Admittedly, there is some uncertainty in our optical depth
estimates due to the rather rough sampling of the cluster
population, especially in cluster redshift. This may contribute to some
of the discrepancies mentioned above. However, as our results
deviate from those of \cite{FedeliEtal2008} and
\citet{LiEtal2005} in opposite directions, it is certainly not the only
cause. Also the relative changes in the optical depths due to
including secondary structures along the line-of-sight and the stellar
components of galaxies should be robust as sparse sampling will affect
all results in a similar way.

\subsection{Einstein radii}
\label{sec:Einstein_radii}

It is straightforward to define an Einstein radius for a
spherically symmetric lens, but there are several ways to generalise the concept  of an Einstein radius to realistic, i.e. non-spherical, cluster lenses. In Eq.~\eqref{eq:def_einstein_radius_crit_curve}, we defined the Einstein radius
$\thetaE$ of a cluster via the solid angle enclosed by its outer
critical curve. Another common definition is the radius $\theta'_{\rm
E}$ of a circular region around the cluster centre having a mean projected surface mass density
equal to the critical density for strong lensing. In the following, we
compare the two definitions.

We computed $\thetaE'$, which we conveniently obtain from each
cluster's projected surface mass density profile, for the 750 most
massive clusters in the Millennium simulation. Furthermore we used
\textit{single-plane} ray tracing \footnote{
We are most interested in large Einstein radii produced by massive
clusters, which are not strongly affected by additional structures along
the line-of-sight.
}
to calculate $\thetaE$ for the
100 most massive Millennium halos and all additional lower mass clusters
available in the cluster sample that we used in the previous sections.
Comparing the results for all clusters for which both values were
derived, we find that the Einstein radii defined via the critical curve
are on average $10-30\%$ larger (depending on cluster mass)
than the radii obtained from the surface density profiles. For example,
the mean and $1\sigma$-scatter of the ratios $\thetaE/\thetaE'$ for
sources at $\zS$=2.1 and clusters at redshift $\zL$=0.28 in mass bins
(a)-(e) are $1.12\pm0.12$, $1.15\pm0.18$, $1.18\pm0.18$,
$1.32\pm0.11$, and 1.28, respectively. \footnote{No scatter is given for
mass bin (e) as it contains only one object.} The scatter is mostly
uniform with only very few large outliers, which are related to merger
processes.

\begin{figure}
\centerline{\includegraphics[width=\linewidth]{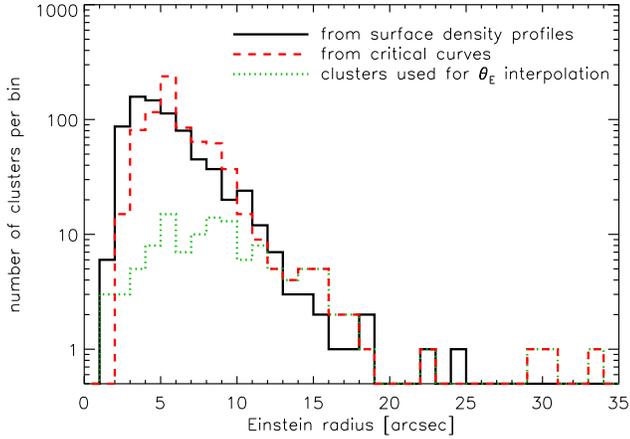}}
\caption{
\label{fig:einstein_dist}
Distribution of Einstein radii of the 750 most massive
Millennium clusters at $\zL$=0.28 and sources at $\zS=2.1$. Distributions are given for the
Einstein radius $\thetaE'$, derived from surface density profiles
(\textit{solid line}), and $\thetaE$, derived from the critical
curves (\textit{dashed line}). Also shown is the $\theta_{\rm
E}$-distribution of the clusters the $\thetaE$-interpolation was
based on (\textit{dotted line}).}
\end{figure}

The difference between $\thetaE'$ and $\thetaE$ can also
be seen in Fig.~\ref{fig:einstein_dist}, which shows the distribution
of Einstein radii $\thetaE'$ and $\thetaE$ of the 750 most
massive clusters in the Millennium simulation at $\zL$=0.28, i.e.
for all clusters at that redshift with $\Mcluster>
1.47\times10^{14}h^{-1}\,\Msolar$. The $\thetaE$-distribution is
based on 120 clusters for which $\thetaE$ was calculated
employing \textit{single-plane} ray tracing. All other clusters were
assigned a $\thetaE$-value by linear interpolation in cluster
mass. Obviously, the $\thetaE$-distribution is shifted towards
larger values compared to the $\thetaE'$-distribution. The
Einstein radius distribution functions shown here should also be
valuable for calibrating semi-analytical model predictions of the
Einstein radius distribution such as those in \cite{OguriBlandford2009},
who suggested they might be used as cosmological probes.

\subsection{Radial distribution of giant arcs}
\label{sec:arc_radii}

Here, we study at what cluster-centric radii long thin arcs are most
likely to be found. For a perfectly
spherical symmetric halo, the longest arcs are expected to appear close
to the Einstein radius. However, for a realistic cluster, this is, in
general, not the case. Even prominent giant arcs that consist of
multiple
merged images of a background galaxy and thus cross the tangential
critical curve may be significantly outside the Einstein radius due to
the critical curve's ellipticity. Arcs that do not consist of multiple
images can be found at even larger radii. This can be seen in the left
panels of Fig.~\ref{fig:crit_curves_caustics}, which show that the
centres of arcs with length to width ratios exceeding 7.5 are typically
located outside the most distant parts of the tangential critical
curve. 

\begin{figure}
\centerline{\includegraphics[width=\linewidth]{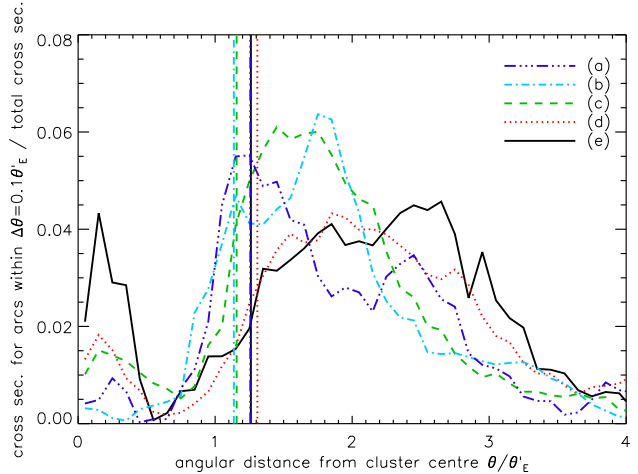}}
\caption{
Radial distribution of arcs with $L/W>7.5$ for sources at redshift $\zS=2.1$ and clusters in mass
bins (a)-(e) at redshift $\zL=0.28$.
The angular distance is given in units of the Einstein-radius
$\thetaE'$, which is calculated for each cluster from a projected
surface mass density profile. The vertical lines indicate the
Einstein-radii derived from the areas enclosed by the clusters'
critical curves. \textit{Full} ray-tracing was used.}
\label{fig:rad_profile}
\end{figure}

We investigate the radial distribution of arcs in more detail by
calculating cross-sections for arcs with $L/W>7.5$ as a function of radial distance
in specific radial bins. In Fig.~\ref{fig:rad_profile}, the results are
shown for the five cluster mass bins listed in Table
\ref{tab:mass_bins}. In total 39 clusters at redshift $\zL=0.28$
and \textit{full} ray-tracing were used in this analysis. The
cluster-centric angular distance is given in units of the Einstein
radius $\thetaE'$. All curves were normalized by the total strong-lensing cross-section for arcs at all radii. Also indicated are the mean
Einstein radii obtained from the critical curves of the clusters in each
bin.\footnote{The indicated $\thetaE/\thetaE'$ values
differ slightly from the numerical values quoted previously in the text
as \textit{full} ray-tracing and a smaller cluster sample were used
here.}

One can clearly distinguish the contribution of radial and tangential
arcs to the radial distribution of arcs. The former are typically found
at $\theta < 0.6\,\thetaE'$. In contrast, the radial distribution of
tangential arcs is very broad, and extends from $\sim 0.6\,\thetaE'$ out
to $\sim 4\,\thetaE'$. Using a larger threshold value for
the length-to-width ratio makes the distribution only slightly narrower. 
For $L/W>10$, it still extends to values $\gtrsim 3\,\thetaE'$. This shows that individual tangential
arcs are not well suited for constraining Einstein radii, instead more
detailed cluster mass models are needed. It is, thus, important to keep
the broad distribution of tangential arcs in cluster-centric radius in
mind when interpreting strong-lensing observations of galaxy clusters.

\section{Summary and conclusions}
\label{sec:summary}

We have used ray-tracing through the Millennium simulation to study how
secondary matter
structures along the line-of-sight affect strong cluster lensing, in particular the cross-section for giant arcs.
We also investigated the impact of the stellar components of
both cluster and interloping galaxies on cluster lensing efficiencies.
Furthermore, we explored the distribution of the cluster Einstein radii and the radial 
distribution of giant arcs. 

We performed ray-tracing along backward light cones using a
multiple-lens-plane algorithm that takes light deflection by all matter
structures between source and observer into account. We also performed
single-lens-plane ray-tracing simulations that either considered only
light deflection by the cluster itself or by all the matter contained in
the same lens plane as the cluster. Thus in the last case, the effects
of correlated structures were included while independent matter
structures at different redshifts were not.

Comparing the results obtained by the different kinds of ray-tracing allows us to clearly pin down the impact
of additional matter structures along the line-of-sight:
\begin{itemize}
 \item The strong-lensing efficiency of clusters increases when including the effects of additional structures along the line-of-sight.
 \item The enhancement is larger for higher source redshifts.
 \item The enhancement is mainly due to structures along the line-of-sight that are not correlated with the lens.
 \item For individual clusters, boosts of the strong-lensing cross-section of up to $50\%$ occur frequently for sources at redshift 3 and higher.
 \item The strong cluster lensing optical depth increases by $10\%$ for
$\zS=2.1$ and by $25\%$ for $\zS=5.7$ when including the effects of
additional structures along the line-of-sight.
\end{itemize}

Using the Millennium simulation's semi-analytic galaxy catalogue, we also investigated how the stellar components of
galaxies affect the lensing properties of clusters. Comparing the results obtained when neglecting and including light deflection by the stellar components of cluster and interloping galaxies, we find:
\begin{itemize}
 \item The strong-lensing properties of massive clusters are only mildly affected by stellar matter.
 \item The lensing cross-sections of less massive clusters can be significantly boosted by cluster galaxies.
 \item The enhancement due to the stellar matter is largest for low cluster and source redshift. 
 \item Interloping galaxies do not affect the cross-section significantly.
 \item Including stellar matter boosts the strong-lensing optical depth
by a factor of 2 for $\zS=1$ and by $25\%$ for $\zS=5.7$.
\end{itemize}

Together, the stars and the additional matter along the line-of-sight
increase the strong-lensing optical depth by a factor of 1.5 to 2,
depending on source redshift.

Furthermore, we computed the distribution of Einstein radii using two different definitions of the Einstein radius.
We also determined the radial distribution of arcs compared to the Einstein radius
of the lensing cluster. We obtain the following results:
\begin{itemize}
 \item The Einstein radii defined via surface mass density profiles and via the area enclosed by the critical curve differ by $10\%$ to $30\%$.
 \item The contributions of radial and tangential arcs to the distribution of arcs in cluster-centric radius can be clearly distinguished.
 \item The radial distribution of tangential arcs is very broad and
extends out to several Einstein radii. Thus, individual arcs are not
well suited for constraining Einstein radii, instead, more detailed mass
models are needed.
\end{itemize}

The work presented here shows that both additional structures along the
line-of-sight and stellar matter in cluster galaxies do affect the
strong-lensing properties of galaxy clusters significantly. Since these
effects can boost cluster strong-lensing cross-sections by factors
$\gtrsim 2$, they need to be taken into account when using cosmological
simulations to derive predictions for the giant arc abundance and
comparing them to observations. Just by themselves, however, they seem
to be too small to fully account for the reported discrepancy between
predicted and observed giant arc abundance in a low-$\sigma_8$ universe
\citep[see, e.g.,][]{FedeliEtal2008}. Thus, other ingredients appear to
be needed in addition to solve the arc statistics problem.

An obvious extension to the work presented here is the calculation of the expected number of giant arcs in surveys with various selection functions. This will require an accurate knowledge of the source redshift and luminosity distribution, in particular at high redshifts, as well as a proper treatment of magnification bias. 

The discrepancies between our results for the optical depths and those based on lower-resolution simulations clearly show the need for structure formation simulations that accurately resolve the inner regions of clusters on scales of a few kpc. On these scales, the complex interplay between dark matter, gas, and stars is expected to strongly influence the matter distribution. Advances in structure formation simulations that incorporate physical processes such as cooling, star formation, and feedback will help to quantify the impact of baryonic physics on the giant arc abundance more accurately.

\section*{Acknowledgments}
We would like to thank Simon White, Matthias Bartelmann, and Peter Schneider for many
helpful discussions. We are also indebted to Massimo Meneghetti
for access to a previous version of the code used for placing sources
and finding the parameters of their images. Furthermore, we would
like to express our gratitude to Volker Springel, Simon White and
Gabriella De Lucia for providing us access to the Millennium simulation
and the semi-analytic galaxy catalogue. This work was supported in parts
by the DFG within the Priority Programme 1177 under the projects SCHN
342/6 and WH 6/3.

\appendix

\bibliographystyle{mnbst}
\bibliography{Astro}
\end{document}